\documentclass[prl,aps,twocolumn,superscriptaddress,showpacs,preprintnumbers,
               amsmath,amssymb,floatfix]{revtex4}

\def\gtwid{\mathrel{\raise.3ex\hbox{$>$\kern-.75em\lower1ex\hbox{$\sim$}}}}
\def\ltwid{\mathrel{\raise.3ex\hbox{$<$\kern-.75em\lower1ex\hbox{$\sim$}}}}

\usepackage{epsfig}

\def\gev{GeV~c$^{-2}$}

\begin{document}

\title{New Results from the Cryogenic Dark Matter Search
Experiment}


\affiliation{Department of Physics, Brown University,
Providence, RI 02912, USA}
\affiliation{Department of Physics, Case Western
Reserve University, Cleveland, OH  44106, USA}
\affiliation{Fermi National Accelerator Laboratory,
Batavia, IL 60510, USA}
\affiliation{Lawrence Berkeley National Laboratory,
Berkeley, CA 94720, USA}
\affiliation{National Institute of Standards and
Technology, Boulder, CO 80303, USA}
\affiliation{Department of Physics, Princeton
University, Princeton, NJ 08544, USA}
\affiliation{Department of Physics, Santa Clara University, Santa
Clara, CA 95053, USA}
\affiliation{School of Physics \& Astronomy, University of Minnesota,
Minneapolis, MN 55455, USA}
\affiliation{Department of Physics, Stanford University,
Stanford, CA 94305, USA}
\affiliation{Department of Physics, University of
California, Berkeley, CA 94720, USA}
\affiliation{Department of Physics, University of
California, Santa Barbara, CA 93106, USA}
\affiliation{Department of Physics, University of
Colorado at Denver, Denver, CO 80217, USA}
\author{D.S.~Akerib} \affiliation{Department of Physics, Case Western
Reserve University, Cleveland, OH  44106, USA}
\author{J.~Alvaro-Dean} \affiliation{Department of Physics, University of
California, Berkeley, CA 94720, USA}
\author{M.S.~Armel} \affiliation{Department of Physics, University of
California, Berkeley, CA 94720, USA}
\author{M.J.~Attisha} \affiliation{Department of Physics, Brown University,
Providence, RI 02912, USA}
\author{L.~Baudis} \affiliation{Department of Physics, Stanford University,
Stanford, CA 94305, USA}
\author{D.A.~Bauer} \affiliation{Department of Physics, University of
California, Santa Barbara, CA 93106, USA}
\author{A.I.~Bolozdynya} \affiliation{Department of Physics, Case Western
Reserve University, Cleveland, OH  44106, USA}
\author{P.L.~Brink} \affiliation{Department of Physics, Stanford
University, Stanford, CA 94305, USA}
\author{R.~Bunker} \affiliation{Department of Physics, University of
California, Santa Barbara, CA 93106, USA}
\author{B.~Cabrera} \affiliation{Department of Physics, Stanford
University, Stanford, CA 94305, USA}
\author{D.O.~Caldwell} \affiliation{Department of Physics, University of
California, Santa Barbara, CA 93106, USA}
\author{J.P.~Castle} \affiliation{Department of Physics, Stanford
University, Stanford, CA 94305, USA}
\author{C.L.~Chang} \affiliation{Department of Physics, Stanford
University, Stanford, CA 94305, USA}
\author{R.M.~Clarke} \affiliation{Department of Physics, Stanford
University, Stanford, CA 94305, USA}
\author{M.B.~Crisler} \affiliation{Fermi National Accelerator Laboratory,
Batavia, IL 60510, USA}
\author{P.~Cushman} \affiliation{School of Physics \& Astronomy, University
of Minnesota, Minneapolis, MN 55455, USA}
\author{A.K.~Davies} \affiliation{Department of Physics, Stanford
University, Stanford, CA 94305, USA}
\author{R.~Dixon} \affiliation{Fermi National Accelerator Laboratory,
Batavia, IL 60510, USA}
\author{D.D.~Driscoll} \affiliation{Department of Physics, Case Western
Reserve University, Cleveland, OH  44106, USA}
\author{L.~Duong} \affiliation{School of Physics \& Astronomy, University of
Minnesota, Minneapolis, MN 55455, USA}
\author{J.~Emes} \affiliation{Lawrence Berkeley National Laboratory,
Berkeley, CA 94720, USA}
\author{R.~Ferril} \affiliation{Department of Physics, University of
California, Santa Barbara, CA 93106, USA}
\author{R.J.~Gaitskell} \affiliation{Department of Physics, Brown
University, Providence, RI 02912, USA}
\author{S.R.~Golwala} \affiliation{Department of Physics, University of
California, Berkeley, CA 94720, USA}
\author{M.~Haldeman} \affiliation{Fermi National Accelerator Laboratory,
Batavia, IL 60510, USA}
\author{J.~Hellmig} \affiliation{Department of Physics, University of
California, Berkeley, CA 94720, USA}
\author{M.~Hennessey} \affiliation{Department of Physics, Stanford
University, Stanford, CA 94305, USA}
\author{D.~Holmgren} \affiliation{Fermi National Accelerator Laboratory,
Batavia, IL 60510, USA}
\author{M.E.~Huber} \affiliation{Department of Physics, University of
Colorado at Denver, Denver, CO 80217, USA}
\author{S.~Kamat} \affiliation{Department of Physics, Case Western Reserve
University, Cleveland, OH  44106, USA}
\author{M.~Kurylowicz} \affiliation{Department of Physics, Stanford
University, Stanford, CA 94305, USA}
\author{A.~Lu} \affiliation{Department of Physics, University of
California, Berkeley, CA 94720, USA}
\author{R.~Mahapatra} \affiliation{Department of Physics, University of
California, Santa Barbara, CA 93106, USA}
\author{V.~Mandic} \affiliation{Department of Physics, University of
California, Berkeley, CA 94720, USA}
\author{J.M.~Martinis} \affiliation{National Institute of Standards and
Technology, Boulder, CO 80303, USA}
\author{P.~Meunier} \affiliation{Department of Physics, University of
California, Berkeley, CA 94720, USA}
\author{N.~Mirabolfathi} \affiliation{Department of Physics, University of
California, Berkeley, CA 94720, USA}
\author{S.W.~Nam} \affiliation{Department of Physics, Stanford University,
Stanford, CA 94305, USA}
\author{H.~Nelson} \affiliation{Department of Physics, University of
California, Santa Barbara, CA 93106, USA}
\author{R.~Nelson} \affiliation{Department of Physics, University of
California, Santa Barbara, CA 93106, USA}
\author{R.W.~Ogburn} \affiliation{Department of Physics, Stanford University,
Stanford, CA 94305, USA}
\author{J.~Perales} \affiliation{Department of Physics, Stanford
University, Stanford, CA 94305, USA}
\author{T.A.~Perera} \affiliation{Department of Physics, Case Western
Reserve University, Cleveland, OH  44106, USA}
\author{M.C.~Perillo Isaac} \affiliation{Department of Physics, University
of California, Berkeley, CA 94720, USA}
\author{W.~Rau} \affiliation{Department of Physics, University of
California, Berkeley, CA 94720, USA}
\author{A.~Reisetter} \affiliation{School of Physics \& Astronomy,
University of Minnesota, Minneapolis, MN 55455, USA}
\author{R.R.~Ross} \affiliation{Lawrence Berkeley National Laboratory,
Berkeley, CA 94720, USA} \affiliation{Department of Physics, University of
California, Berkeley, CA 94720, USA}
\author{T.~Saab} \affiliation{Department of Physics, Stanford University,
Stanford, CA 94305, USA}
\author{B.~Sadoulet} \affiliation{Lawrence Berkeley National Laboratory,
Berkeley, CA 94720, USA} \affiliation{Department of Physics, University of
California, Berkeley, CA 94720, USA}
\author{J.~Sander} \affiliation{Department of Physics, University of
California, Santa Barbara, CA 93106, USA}
\author{C.~Savage} \affiliation{Department of Physics, University of
California, Santa Barbara, CA 93106, USA}
\author{R.W.~Schnee} \affiliation{Department of Physics, Case Western
Reserve University, Cleveland, OH  44106, USA}
\author{D.N.~Seitz} \affiliation{Department of Physics, University of
California, Berkeley, CA 94720, USA}
\author{T.A.~Shutt} \affiliation{Department of Physics, Princeton
University, Princeton, NJ 08544, USA}
\author{G.~Smith} \affiliation{Department of Physics, University of
California, Berkeley, CA 94720, USA}
\author{A.L.~Spadafora} \affiliation{Department of Physics, University of
California, Berkeley, CA 94720, USA}
\author{J-P.F.~Thompson} \affiliation{Department of Physics, Brown
University, Providence, RI 02912, USA}
\author{A.~Tomada} \affiliation{Department of Physics, Stanford University,
Stanford, CA 94305, USA}
\author{G.~Wang} \affiliation{Department of Physics, Case Western Reserve
University, Cleveland, OH  44106, USA}
\author{S.~Yellin} \affiliation{Department of Physics, University of
California, Santa Barbara, CA 93106, USA}
\author{B.A.~Young} \affiliation{Department of Physics, Santa Clara
University, Santa Clara, CA 95053, USA}

\collaboration{CDMS Collaboration}

\noaffiliation

\date{\today}

\begin{abstract}

Using improved Ge and Si detectors, 
better neutron shielding, 
and increased counting time, 
the Cryogenic Dark Matter Search
(CDMS) experiment has obtained stricter limits
on the cross section of weakly interacting massive
particles (WIMPs) elastically scattering from nuclei.  Increased
discrimination against electromagnetic backgrounds and reduction of the
neutron flux confirm 
WIMP-candidate events previously detected by CDMS were 
consistent with
neutrons and give limits on spin-independent WIMP interactions which
are $>2\times$ lower 
than previous CDMS results for high WIMP mass, and
which exclude new parameter space for WIMPs with mass between 8--20\ \gev.

\end{abstract}


\maketitle

This Letter reports new exclusion limits from
the Cryogenic Dark Matter Search
(CDMS) experiment on the wide class of nonluminous, nonbaryonic, 
weakly interacting massive
particles (WIMPs)~\cite{lee,jkg} which could 
constitute most of the matter in the universe~\cite{bergstrom}. 
A natural WIMP candidate is provided by supersymmetry in the form of the
stable lightest superpartner, 
usually taken to be a neutralino of typical
mass $\sim100$ GeV/c$^2$~\cite{jkg,ellis}.  
Since the WIMPs are expected to be in a roughly isothermal halo within
which the visible portion of our galaxy resides, the energy given to a
Ge or Si detector nucleus scattered elastically 
by a WIMP would be only a few to tens of keV~\cite{lewin}.

Because of this low recoil energy and very low event rate 
($<1$ event per day per kg of detector mass),
it is essential to suppress backgrounds drastically.
The CDMS detectors 
discriminate nuclear recoils (such as would be produced
by WIMPs) from electron recoils by measuring both ionization and phonon
energy, greatly reducing the otherwise dominant electromagnetic background.
The ionization is much less for nuclear than for electron recoils,
while the phonon signal enables a determination of
the recoil energy.  The main
remaining background is from neutrons, which produce WIMP-like recoils, and
hence must be distinguished by other means.  Two are employed: 1)
while Ge and Si have similar scattering rates per nucleon for neutrons, Ge
is 5--7 times more efficient than Si for coherently scattering WIMPs; 
2) a single WIMP will not scatter in more than one detector, while a neutron 
frequently will.

While brief reviews of all parts of the experiment 
are provided below, most
details have been published~\cite{r19prd}, and therefore the emphasis here
will be on the differences from previous work.  
The previously published results are from three 
165 g Ge BLIP (Berkeley Large Ionization-and-Phonon-mediated) and one 100 g
Si ZIP (Z-sensitive Ionization and Phonon-mediated)
detectors.
The latter, employed as one measure of background neutrons, was not used
simultaneously with the Ge BLIPs, but rather in a separate run.  
BLIP detectors determine phonon production 
from the detector's calorimetric temperature change, whereas 
ZIP detectors~\cite{zips} collect 
athermal phonons 
to provide both phonon production and position information.
Position information can be obtained from pulse arrival times and
relative signal sizes in multiple sensors. 
Pulse rise time gives further discrimination against
surface events which could otherwise be misidentified as nuclear recoils.  
The new experiment used four 250 g Ge
and two 100 g Si ZIP detectors simultaneously, improving 
measurement of neutron backgrounds.

Since new detector technologies were used, both CDMS
experiments operated at a convenient shallow site. 
The experiments had an overburden of 16 meters water equivalent, sufficient
to stop the hadronic cosmic ray component and reduce the muon flux by a
factor of 5.
The apparatus from outside to inside included
a plastic-scintillator veto to reject muon-coincident particles
by a factor of $>10^3$, a 15-cm-thick lead shield to reduce 
background 
photon flux by a factor of $10^3$, 25 cm of
polyethylene, a 20 mK volume provided by
a custom, radiopure side extension to an
Oxford 400S dilution refrigerator \cite{ref:8a}, and
1 cm of ancient Pb.  In the 
new experiment 11 kg of polyethylene added inside the ancient Pb
reduced the neutrons by a 
factor of 2.3, not only making this a better
measurement, but also 
confirming 
the neutron level given in the previously published experiment.

Because detector discrimination reduced
electromagnetic backgrounds, 
the limiting background was from relatively rare,
high-energy
neutrons produced outside the veto. 
The neutrons may ``punch through" the shielding and yield
secondary neutrons 
whose scatters in the detectors can
mimic the WIMP signal~\cite{r19prd}.
As in the previous experiment, the propagation of these neutrons was
simulated accurately, as confirmed by comparison with veto-coincident and
calibration-source neutrons.  There is excellent agreement of the simulated
and observed recoil-energy spectra. 
No experimental results depend on any {\em a priori}\/ knowledge
of the absolute value of the neutron background.

The utility of the simulations lies in predicting 
the relative rates of neutron events in the Ge and Si detectors, and the
relative number of neutron scatters in a single detector to 
that in multiple detectors.  
These predicted ratios and the observed numbers 
of multiple-detector and Si neutrons
determine the single-scatter neutron background in the Ge detectors.  
A check on the
consistency of the null hypothesis that all candidate events in
the previous and present experiments 
are neutrons is provided by comparing the numbers of observed and
predicted events in the two experiments simultaneously. 
Assuming all events are due to neutrons and taking into account the different
polyethylene shielding, the numbers of Ge
and Si singles and multiples 
observed in the two experiments all agree with the same incident neutron flux. 

The new 1-cm-thick, 7.62-cm-diameter detectors were stacked 3.5 mm apart with
no intervening material.  The ionization electrodes deposited on the 
bottom surface of each detector were divided into an annular outer
part shielding the disk-shaped inner part 
from any low-energy electrons emitted from surrounding surfaces.  
The disk part constituted 85\% of the physical volume in the new experiment
in contrast to $\sim56$\% previously.
The disk areas at only the top and bottom 
of the stack were exposed to external materials. 
A Si detector, 
known to have had some exposure to
$^{14}$C contamination during a 
previous run, was placed on the bottom and
was used for multiple scatters but not for 
singles data.
The Ge detector requiring a 20-keV analysis threshold was put at the top.
All other detectors (from top to bottom: Ge, Ge, Si, Ge)
had analysis thresholds of 5 keV, compared to
the 10-keV thresholds used in the previous
experiment.  The analysis threshold was 
well above the phonon-trigger 
threshold of $\sim$2 keV and the $\sim$1.5 keV (for electron recoils)
search threshold for ionization from the disk region. 

\begin{figure}
\epsfig{figure=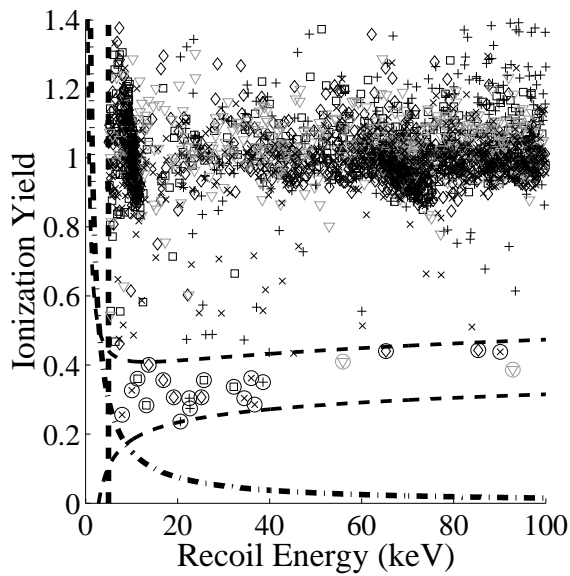, width=7.7cm}
\caption{\label{yplot}
Ionization yield (Y) vs.\ recoil energy for veto-anticoincident single
scatters in the 4 Ge (black $\times$, 
diamond, square, and $+$) and 1 inner Si (grey 
triangle) detectors above the 5 keV analysis threshold (dashed
vertical line).  Twenty events (circled) lie within the mean nominal 95\%
nuclear-recoil acceptance region (dashed curves), above the ionization
threshold (dot-dashed curve).
Most of the events with $0.5<{\rm Y}<0.8$
are in the top Ge detector ($\times$), whose top face is unshielded, or in
the bottom Ge detector ($+$), whose bottom disk region faces the
contaminated Si detector.
}
\end{figure}

Thresholds were 
determined by how well the detector's ionization yield, Y, the ratio of
ionization production to recoil energy (using also the phonon
signal), separates electron from nuclear recoils.  Photons cause most bulk
electron recoils, while low-energy electrons can cause low-Y electron
recoils in a thin surface layer, possibly resulting in confusion
with nuclear recoils.  
In this
experiment ZIP detectors provided additional rejection of such
surface events 
by pulse rise time information, which is sensitive to the depth of the
interaction.  Neutron and photon sources 
were used to determine efficiencies for discrimination between nuclear
recoils and bulk and surface electron events.  These calibration runs
established the acceptable nuclear---and hence WIMP---recoil band.  
Figure~\ref{yplot} illustrates the particle discrimination, 
which can be characterized by a
photon rejection $>99.98$\% (5--100 keV) and an electron rejection $>99$\%
above 10 keV for the four central detectors, far exceeding those (99.9\%
and 95\%, respectively) of the previous experiment.

The data set of the present experiment was taken with 
a 3 V bias
voltage across the detector ionization electrodes used to collect the
electrons and holes.  A data set to be reported later employed a 6 V bias.
The larger voltage improves the ionization yield of surface electron events 
but
results in worse rise-time-based particle discrimination.
At 3 V bias, 93 days of low-background data were taken from 
December 2001 to April 2002,
resulting in 65.8 live days and $4.6\times10^6$ events.  After cuts this
became 28.3 kg-d of data, substantially more than the 15.8~kg-d of the
previous experiment.  Three calibrations with $^{60}$Co photon sources
and two with $^{252}$Cf neutron sources 
were performed at various times during the run.

The position information available from the ZIP detectors was used to make
small corrections for variations of the phonon signal with 
event location.
The corrections for this position dependence,
improving the phonon energy resolution and the surface electron rejection,
were obtained from the photon calibration.

Most cut parameters were set based on calibration data, and all cuts except
the threshold-energy cut for the top detector were set 
before examining the 
final 90\% of the low-background data.  The cut
for data quality had $>99.99$\% efficiency.  Having
at least 80\% of the ionization energy in
the disk part of the detector and having phonon rise time
$>12~\mu$s for Ge and $>6~\mu$s for Si gave an energy-dependent
efficiency for nuclear-recoil events varying from
10--15\% at 5 keV to 40--45\% at 20 keV to 50-60\%
at higher energies.
Requiring $>40~\mu$s after the most recent muon veto gave an $\sim80$\%
efficiency for a typical 5.2 kHz veto rate.  Selecting a
nuclear-recoil
band in Y within $\pm2~\sigma$ of the band mean gave 
95\% efficiency for nuclear-recoil events.

Nuclear-recoil, single-scattering candidates satisfied 
all the above cuts and had energy above both the ionization
and phonon thresholds in one detector, but no phonon
trigger in any other detector within $50~\mu$s of the event trigger.
Nuclear-recoil, multiple-scattering candidates required 
passing data-quality and veto-anticoincident cuts, 
all triggering detectors having 
between 5--100 keV of recoil energy (and at least
80\% of their ionization energy in their disk region), 
at least two of the detectors passing the nuclear-recoil cut, and 
at least one of the detectors passing the rise-time cut.

\begin{figure}
\epsfig{figure=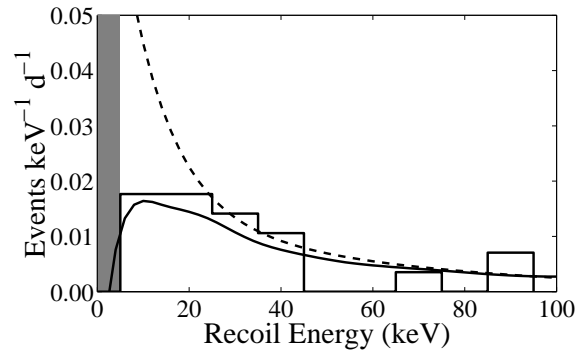, width=7.6cm}
\caption{\label{spec}
Histogram in 10-keV bins of the 20 veto-anticoincident,
single-scatter nuclear-recoil candidates observed in
the 4 Ge 
detectors of total mass 1 kg.  The dashed curve is the shape of the
expected 
recoil-energy spectrum due to incident neutrons, while the solid curve
also takes into account the detection efficiency and is normalized to
20 events.
}
\end{figure}

A histogram of the 20 Ge single-scatter nuclear-recoil candidates as a
function of energy is shown in Fig.~\ref{spec}, along with
the expected neutron spectrum.  This simulated neutron spectrum agrees
well with the data, as is verified by a Kolmogorov-Smirnov test.
These 20 candidates could include some surface-electron events.
The expected number of such events in the nuclear-recoil band is 
$1.2\pm0.3$ in the Ge detectors and $0.8\pm0.6$ in the Si detector.
The expected contamination was found from maximum likelihood fits 
to the rise time and yield (Y) 
distribution of the $\sim1$\% of photon
calibration events occurring near the surface.
The fits were
normalized to the observed number of background surface events
outside the signal region. 

\begin{figure}
\epsfig{figure=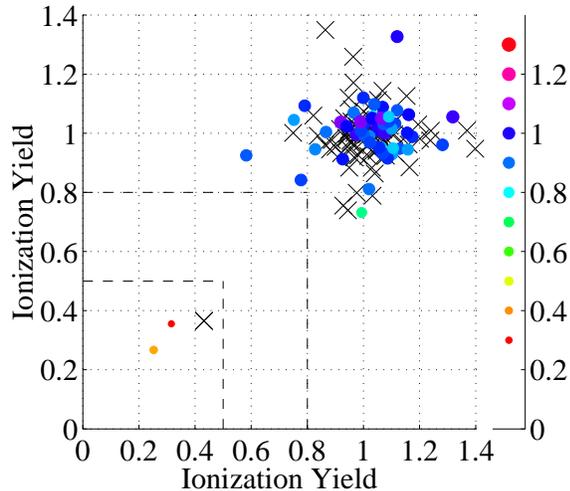, width=7.55cm}
\caption{\label{multiples}
Scatter plot of ionization yields for veto-anticoincident
triple scatters (filled circles) and non-nearest-neighbor double
scatters ($\times$'s)
with all scatters between 5 and 100~keV and within the
fiducial volume.  For triple scatters, the size and color of the circle
indicates the ionization yield in the third detector.
Note both neutron triple-scatter candidates show low yield in all
three detectors, while no other triple scatters have low yield in any
detector.
As expected for such events, there is clearly negligible
contamination
from surface events.}
\end{figure}

The 90\%~CL excluded region for the WIMP mass $M$ and the
spin-independent WIMP-nucleon elastic-scattering cross-section
$\sigma$ is derived using a likelihood ratio 
approach and conservative parameter values as described in~\cite{r19prd}.
This method accounts for those observed Ge single
scatters that are due to the unvetoed neutron flux,
as constrained by the observed
2 triple scatters and 1 non-nearest-neighbor double scatter
(shown in Fig.~\ref{multiples}), along with 2 single-scatter
nuclear-recoil candidates in the Si detector.  Nearest-neighbor
double scattering events were not used to determine the number
of neutrons in the Ge single-scatter sample because correction for
false events due to double scattering of surface electrons is too
uncertain at this time.

\begin{figure}[t]
\epsfig{figure=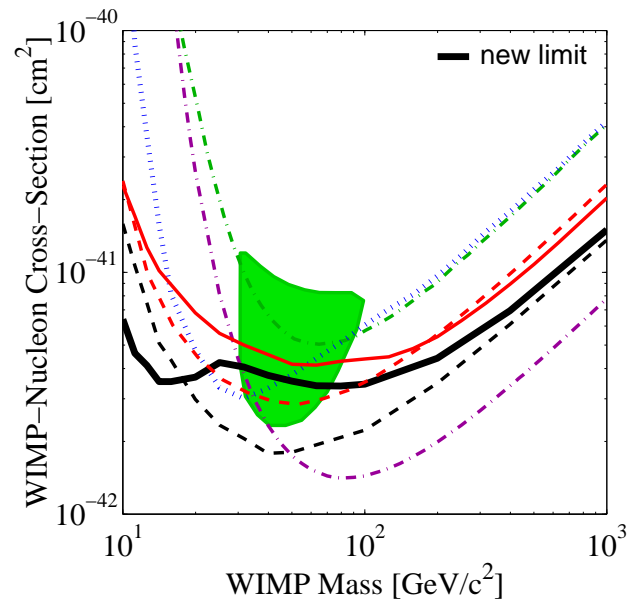, width=8.3cm}
\caption{Spin-independent $\sigma$ vs.\ $M$.  The regions above the
curves are excluded at 90\%~CL.  Solid, thick black curve: limit from this
analysis including statistical subtraction of the neutron
background.
Solid red curve: limit from this
analysis without statistical subtraction of the neutron background.
Dashed curves: CDMS expected sensitivity (median simulated
limit) given the expected neutron background (as normalized based
on this and previous work) of 3.3
multiple-scatters, 18~single scatters in Ge,
and an expected background in Si of 0.8~electrons and 3.6~neutrons,
with (black) or without (red) neutron subtraction.
Blue dotted curve: previous CDMS upper limit~\protect\cite{r19prd}.
Green dot-dashed
curve: DAMA limit using pulse-shape analysis~\protect\cite{DAMApsa}.
The DAMA 3$\sigma$ allowed region not including
the DAMA limit~\protect\cite{DAMA2000} is shown as a shaded region.
Above 30~\gev, the EDELWEISS~\cite{edel2002} (purple 
dot-dashed curve) experiment provides more sensitive limits.
All curves are normalized
following~\protect\cite{lewin}, 
using the Helm spin-independent
form-factor, A$^2$~scaling, WIMP characteristic velocity $v_0 = 220$
km~s$^{-1}$, mean Earth velocity $v_E = 232$~km~s$^{-1}$, and $\rho =
0.3$~GeV~c$^{-2}$~cm$^{-3}$.
}
\label{limitplot}
\end{figure}

Figure~\ref{limitplot} displays the
resulting upper limits on the spin-independent WIMP-nucleon
elastic-scattering cross-section.  Because the neutron
simulation predicts 5.3 Ge singles per multiple-detector neutron,
all the nuclear recoils may be
neutron scatters and $\sigma = 0$ is not excluded.
The dip in the limit curve is due to the fact that low-mass WIMPs
could not cause the several detected events between $\sim$30--40 keV.
Figure~\ref{limitplot} also shows the upper limits calculated
ignoring all information about the neutron background, using
Yellin's ``Optimum Interval'' method~\cite{yellin}.
The limit is essentially set by a region of the energy spectrum with
few events compared to the number expected from the WIMP energy
spectrum.  Because of the weak statistical power gained by estimating
the neutron background based on the small number of multiple-scatter
and Si events, the sensitivity without subtraction of the neutron
background is nearly as good as the sensitivity
including subtraction of the neutron background.
These limits
exclude new parameter space for WIMPs with $8<M<20$ GeV c$^{-2}$.
Furthermore, a simultaneous goodness-of-fit test (as described
in~\cite{r19prd}),
indicates the model-independent
annual-modulation signal of DAMA (as shown in Fig.~2
of~\cite{DAMA2000})
and these CDMS data are incompatible at 99.98\%~CL,
if the WIMP interactions and halo are standard.

This work is supported by the National Science
Foundation under Grant No. PHY-9722414, by the Department of Energy
under contracts DE-AC03-76SF00098, DE-FG03-90ER40569,
DE-FG03-91ER40618, and by Fermilab, operated by the Universities
Research Association, Inc., under Contract No. DE-AC02-76CH03000 with
the Department of Energy.




\gdef\journal#1, #2, #3, #4#5#6#7{      
    #1~{\bf #2}, #3 (#4#5#6#7)}   

\def\apj{\journal Astrophys.\ J., }
\def\app{\journal Astropart.\ Phys., }
\def\baas{\journal Bull.\ Am.\ Astron.\ Soc., }
\def\ejpc{\journal Eur.\ J.\ Phys.\ C., }
\def\nature{\journal Nature, }
\def\nc{\journal Nuovo Cimento, }
\def\nima{\journal Nucl.\ Instr.\ Meth.\ A, }
\def\np{\journal Nucl.\ Phys., }
\def\npps{\journal Nucl.\ Phys.\ (Proc.\ Suppl.), }
\def\pl{\journal Phys.\ Lett., }
\def\prep{\journal Phys.\ Rep., }
\def\pr{\journal Phys.\ Rev., }
\def\prc{\journal Phys.\ Rev.\ C, }
\def\prd{\journal Phys.\ Rev.\ D, }
\def\prl{\journal Phys.\ Rev.\ Lett., }
\def\rsi{\journal Rev. Sci. Instr., }
\def\rpp{\journal Rep.\ Prog.\ Phys., }
\def\sjnp{\journal Sov.\ J.\ Nucl.\ Phys., }
\def\solarphys{\journal Solar Phys., }
\def\jetp{\journal J.\ Exp.\ Theor.\ Phys., }

\end{document}